\documentclass[12pt]{article}
\usepackage{epsf}
\usepackage{amsmath}
\usepackage{graphics}
\usepackage{graphicx}

\renewcommand{\thefootnote}{\#\arabic{footnote}}

\newcommand{\gtrsim}{ \mathop{}_{\textstyle \sim}^{\textstyle >} }
\newcommand{\lesssim}{ \mathop{}_{\textstyle \sim}^{\textstyle <} }

\begin{document}

\setcounter{footnote}{0}
\begin{titlepage}

\begin{center}

\hfill astro-ph/0605481\\
\hfill May 2006\\

\vskip .5in

{\Large \bf
Implication of Dark Energy Parametrizations 
on the Determination of the Curvature of the Universe
}

\vskip .45in

{\large
Kazuhide Ichikawa$^{1}$, Masahiro Kawasaki$^{1}$, Toyokazu Sekiguchi$^{1}$ \\
and Tomo Takahashi$^{1,2}$
}

\vskip .45in

{\em
$^{1}$Institute for Cosmic Ray Research,
University of Tokyo\\
Kashiwa 277-8582, Japan, \\
$^{2}$Department of physics, Saga University, Saga 840-8502, Japan
}

\end{center}

\vskip .4in

\begin{abstract}

  We investigate how the nature of dark energy affects the 
  determination of the curvature of the universe from recent observations.
  For this purpose, 
  we consider the constraints on the matter and dark energy density 
  using observations of type Ia supernovae, baryon acoustic
  oscillation peak and cosmic microwave background with several types of 
  dark energy equation of state.  Although it is
  usually said that the combination of current observations favors a
  flat universe, we found that a relatively large parameter space
  allows the universe to be open for a particular model of dark energy. 
  We also discuss what kind of dark energy model or prior allow a non-flat universe.

\vspace{1cm}
\end{abstract}

\end{titlepage}

\renewcommand{\thepage}{\arabic{page}}
\setcounter{page}{1}
\renewcommand{\thefootnote}{\#\arabic{footnote}}
\renewcommand{\theequation}{\thesection.\arabic{equation}}

\section{Introduction}
\setcounter{equation}{0}

The nature of dark energy is one of the challenging mysteries in
today's science.  Although cosmological observations such as cosmic
microwave background (CMB) is becoming very precise and a lot of
efforts have been made to understand it, we still do not know what the
dark energy is. One can try to
investigate it phenomenologically using the dark energy equation of
state $w_X$ as a parameter which describes the nature of dark energy
and constraining it using various observations. In the early stage of
the study of dark energy, the equation of state for dark energy has
been assumed to be constant in time.  However, many models proposed so
far have time-varying equation of state, thus most of recent
phenomenological works for dark energy accommodate its time dependence
in some way. 

In such works, constraints on the time dependence of the equation of state 
for dark energy is often obtained assuming a flat universe.  The assumption of a
flat universe seems to be reasonable since it is usually said that the
constraint on the curvature of the universe from a combination of
various data sets suggests that the universe is flat. But notice that,
when one discusses the constraint on the curvature of the universe,
the cosmological constant is usually assumed as dark energy. It is
not so obvious that a flat universe is preferred when one assumes
different models for dark energy. 
Similarly, when one considers the constraint on the nature of dark energy such as the
equation of state, a flat universe is assumed in many literatures, but it
is also not so trivial how the curvature of the universe affects the
constraint on the nature of dark energy.

Thus it is important to
study the curvature of the universe without assuming a cosmological
constant for dark energy or the dark energy equation of state without
assuming a flat universe.  In fact, some works along this line have
been done in Refs.~\cite{Crooks:2003pa,Gong:2005de,Ichikawa:2005nb}
(for other recent considerations on degeneracies between the equation of state for dark energy
and the curvature of the universe, we refer to 
Refs.~\cite{Knox:2005hx, Polarski:2005jr, Dick:2006ev, 
Huang:2006er, Knox:2006ux} ).
In
Ref.~\cite{Crooks:2003pa}, it was shown that there is a degeneracy
between the curvature of the universe and the constant equation of
state for dark energy in the CMB angular power spectrum.  Furthermore,
in Ref.~\cite{Gong:2005de,Ichikawa:2005nb}, the constraint on the
curvature of the universe was investigated using a particular
parameterization of dark energy which accommodates the time dependence
in some way. It was shown that, even if we assume a time-varying
equation of state for dark energy, one can obtain an almost flat
universe from the combination of various observational data
sets. However, it should be noticed that the robustness of the
flatness was shown only for some particular parametrizations for dark
energy. 

In this paper, we extend the analysis done in Ref.~\cite{Ichikawa:2005nb}
in order to elucidate the role of the dark energy prior on the measurement
of the curvature of the universe.
Especially, we will show that there exists a type of dark energy time-dependence
which allows significantly larger region of an open universe than that allowed by
the dark energy model adopted in Ref.~\cite{Ichikawa:2005nb} (and in many other works).
It should be emphasized that this holds even if we combine recent observations
which measure cosmological distances to various redshifts.
We consider the observations of type Ia supernovae (SNeIa),
CMB and baryon acoustic oscillation (BAO) peak. For the
CMB constraint, we make use of the shift parameter which well captures
the whole shift of the CMB angular power spectrum.

The structure of this paper is as follows. In the next section, after
we discuss the parametrization of the dark energy equation of
state, we briefly summarize the analysis method adopted in this
paper.  Then, in section \ref{sec:constraint}, we study the constraints on 
the curvature of the universe. The final section is devoted to the
summary of this paper.

\section{Analysis method} \label{sec:method}
In this section, first we summarize the parameterization for the dark
energy equation of state which we are going to use in our analysis.

We consider the following parametrization for the dark energy equation
of state,
\begin{equation}
w_X(z) = 
\begin{cases}
 w_0 + \displaystyle\frac{w_1 - w_0}{z_*} z 
 &( {\rm for}~~ z \le z_*)  \\ \\
 w_1      & ({\rm for}~~ z \ge z_*),   
\end{cases} 
\label{eq:param}
\end{equation}
where we interpolate the value of $w_X$ linearly from the present 
epoch back to the
transition redshift $z_*$. $w_X$ becomes $w_0$ at $z=0$ and $w_1$ at
$z \ge z_*$.  The energy density of dark energy in this parametriztion
is calculated as
\begin{equation}
\rho_X (z)/ \rho_X(z=0) =
\begin{cases}
 \exp [3 \alpha z] (1+z)^{3(1 + w_0 - \alpha)} 
&(  {\rm for}~~ z \le z_*)  \\ \\
 \exp [3 \alpha z_*] (1+z_*)^{3(1 + w_0 - \alpha)} 
\displaystyle\left( 
\frac{1+z}{1+z_*}\right)^{3(1+w_1)}
& ({\rm for}~~ z \ge z_*),   
\end{cases} 
\end{equation}
where 
\begin{eqnarray}
\alpha & \equiv & \frac{w_1 - w_0}{z_*}.
\end{eqnarray}
This parametrization is essentially the same as the one which has a
linear dependence on the redshift $z$ such as $w_X = w_0 + \alpha
z$. In fact, this form is adopted in many literatures with a cutoff at
some redshift to avoid large value of $w_X$ at $z \sim 1000$ which is
relevant to the CMB constraint
\cite{Huterer:2000mj,Weller:2001gf,Frampton:2002vv}. 
In this paper, we consider several value of $z_*$ and show how the 
values of $z_*$ affect the constraint on the curvature.

Now we briefly discuss the analysis method for constraining the model
and cosmological parameters.  For the purpose of the present paper, we
make use of observational data of SNeIa, CMB and baryon acoustic
oscillation \footnote{
To obtain the constraint on the nature of dark energy, 
some authors also make use of the angular size of compact radio 
sources and/or X-ray gas mass fraction of galaxy cluster 
Ref.\cite{Lima:2001zz,Lima:2003dd,Rapetti:2004aa,Zhu:2004cu}.
However we do not consider them here.
}. To compare a model with observational data,
first we need to know the evolution of the Hubble parameter which can
be written as
\begin{equation}
H^2(z) = H_0^2 \left[ 
\Omega_r (1+z)^4 + \Omega_m(1+z)^3  + \Omega_k(1+z)^2 
+ \Omega_X \exp \left( 
3 \int_0^z ( 1 + w_X(\bar{z})) \frac{d\bar{z}}{1+\bar{z}} \right) 
\right] 
\end{equation}
where $H_0$ is the Hubble parameter at the present time, $\Omega_i$ is
the energy density at the present epoch normalized by the critical
energy density and the subscripts $r,m,k$ and $X$ represent
radiation, matter, the curvature and dark energy, respectively. 

For the SNeIa data, we fit the distance modulus to observational data
from the gold data set given in Ref.~\cite{Riess:2004nr} and the
Supernova Legacy survey \cite{Astier:2005qq}. The distance modulus is
defined as
\begin{equation}
M -m = 5\log d_L + 25. 
\end{equation}
Here $d_L$ is the luminosity distance in units of Mpc which is written as
\begin{equation}
d_L = \frac{1+z}{\sqrt{|\Omega_k|} }
\mathcal{S}  \left( \sqrt{|\Omega_k|} \int_0^z \frac{dz'}{H(z') /H_0} \right)
\end{equation}
where $\mathcal{S}$ is defined as $\mathcal{S}(x) = \sin (x)$ for a
closed universe, $\mathcal{S}(x) = \sinh (x)$ for an open universe and
$\mathcal{S}(x) = x$ with the factor $\sqrt{|\Omega_k|}$ being removed
for a flat universe.

We also use the observation of baryon oscillation acoustic peak which
has been detected in the SDSS luminous red galaxy sample and
recently used to constrain the properties of dark energy by
many authors. The quantity we use to constrain the cosmological
parameters is the parameter $A$ which is defined as
\begin{equation}
A = \frac{\sqrt{\Omega_m}}{(H(z_1)/H_0)^{1/3}} \left[  
\frac{1}{z_1 \sqrt{|\Omega_k|} }
\mathcal{S}  \left( \sqrt{|\Omega_k|} \int_0^{z_1} \frac{dz'}{H(z')/H_0} \right)
\right]^{2/3}
\end{equation} 
where $z_1=0.35$ and $A$ is measured as $A=0.469\pm 0.017$
\cite{Eisenstein:2005su}.

To obtain the constraint from CMB observations, we make use of the
so-called shift parameter which measures the whole shift of the CMB
angular power spectrum. The shift parameter is usually written as $R$
and defined as
\begin{equation}
R  = \frac{\Omega_m}{\sqrt{|\Omega_k|} }
\mathcal{S}  \left( \sqrt{|\Omega_k|} \int_0^{z_2} \frac{dz'}{H(z')/H_0} \right)
\end{equation}
where $z_2=1089$, the redshift of the epoch of the recombination. 
From the three-year WMAP result 
\cite{Jarosik:2006ib,Hinshaw:2006ia,Page:2006hz,Spergel:2006hy},
the shift parameter is constrained to be $R = 1.70 \pm 0.03$
\cite{Wang:2006ts}.  Since we only consider the shift parameter which
is determined only by the background evolution for the constraint from
CMB, we do not need to include the effect of the fluctuation of dark
energy. In this paper, by using shift parameter, we can confine ourselves to 
considering the effects of the modification of the background evolution alone. 

Here, some comments on this are in order.  
The reason why we only consider the background evolution is
twofold.  The first one is that the constraint on dark energy almost
comes from the fit to the position of the acoustic peaks although the
fluctuation of dark energy component can affect the low multipole
region of the CMB angular power spectrum. As mentioned above, the shift
parameter is the measure of the position of the acoustic peaks. Thus
the use of the shift parameter is a robust way to obtain the
constraint from CMB. 
Another reason is that we need to know more about
the nature of dark energy to include the effects of the fluctuation of
it such as its speed of sound.  Furthermore, in some models, the
Friedmann equation itself is modified to explain the present cosmic
acceleration such as the DGP model \cite{Dvali:2000hr}.  Even in such
models, one can define the effective equation of state for a
``fictitious'' dark energy component such that the background
evolution is the same as the original models. However, as for the
perturbation, we need to take the modification into account properly,
which cannot be done by just assuming a ``fictitious'' dark energy in
most cases. 
By considering the observable which is only affected by the modification
of background we can make our analysis general, 
accomodating both ordinary dark energy and modified gravity.

Using the observations mentioned above, we perform $\chi^2$ analysis 
to 
obtain the constraints on 
energy densities of matter and dark energy with its equation of state
which varies as Eq.~(\ref{eq:param}).
In the next section, we discuss the constraints on them in detail.

\section{Constraints on the curvature of the universe} \label{sec:constraint}
In this section, we discuss the constraint on the curvature of the universe
 using observations mentioned in the previous
section.  We assume the parametrization introduced in the previous
section for the equation of state of dark energy.

\begin{figure}[t]
\begin{center}
\scalebox{0.6}{\includegraphics{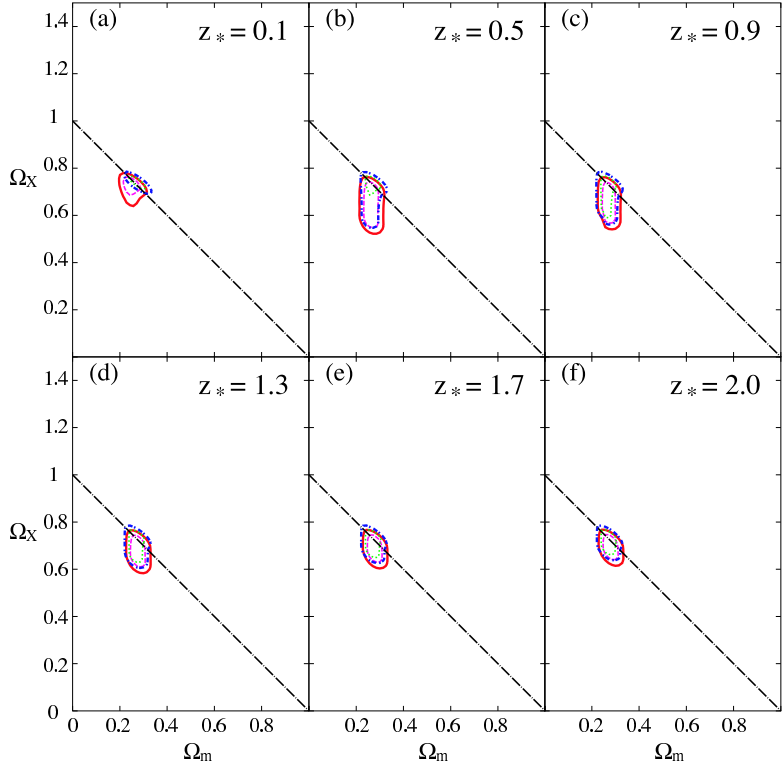}}
\caption{Constraints from all combined data set
on the $\Omega_m$ vs.  $\Omega_X$ plane for the parametrization of
  Eq.~(\ref{eq:param}). We marginalized over $w_0$ and $w_1$.
 1$\sigma$ and 2$\sigma$ contours using gold (SNLS) data set for SNeIa
  are shown with magenta dashed (green dotted) and red solid (blue dot-dashed) 
  lines respectively.
 We fixed the value of the transition redshift
  as $z_*=$0.1 (a), 0.5 (b), 0.9 (c), 1.3 (d), 1.5 (e) and 2.0 (f).
}
\label{fig:OmOX}
\end{center}
\end{figure}

Now we present the constraint on the $\Omega_m$ vs. $\Omega_X$ plane.
In Fig.~\ref{fig:OmOX}, contours of 1$\sigma$ and 2$\sigma$
constraints from all data combined (i.e., SNeIa, CMB and BAO) are
shown. In the figure, we fixed
the value of the transition redshift $z_*$ as $z_* = 0.1, 0.5, 0.9,
1.3, 1.7$ and $2.0$.  To obtain the constraint, we marginalized over
the values of $w_0$ and $w_1$.  We assumed the prior on
$w_0$ and $w_1$ as $ -5 < w_0 < 0$ and $ -5 < w_1 < 0$.
 We note that $\chi^2$ minimum is not given by the lower bound of 
 these priors for any $(\Omega_m, \Omega_X)$. 
 As seen from
the figure, for several values of the transition redshift $z_*$, the
allowed region extends to the parameter space where the curvature of
the universe is negative, namely the universe is open.  
This seems contrary to the result of Ref.~\cite{Ichikawa:2005nb}
where we have shown the same combined data set 
(to be exact, since we used the WMAP 1st year value for the shift parameter in 
Ref.~\cite{Ichikawa:2005nb}, its error was two times as large as the three-year 
value we adopt in this paper) favors the region closely around
a flat universe (see Fig.~5d in Ref.~\cite{Ichikawa:2005nb}) for the dark energy model
\begin{equation}
w_X  = \tilde{w}_0 + (1-a) \tilde{w}_1,
\label{eq:a}
\end{equation}
upon marginalizing over $\tilde{w}_0$ and $\tilde{w}_1$.
In the following, we will give explanations to our result paying particular 
attention to the difference between the models examined in 
Ref.~\cite{Ichikawa:2005nb} and Eq.~(\ref{eq:param}).

Before interpreting the results in Fig.~\ref{fig:OmOX},
we should note that although the model defined by Eq. (\ref{eq:param})
appears to have three parameters $w_0, w_1$ and $z_*$ we regard it as
a family of two-parameter ($w_0$ and $w_1$) models labeled by $z_*$.
Namely, we consider several two-parameter models each with different value of $z_*$. 
Such distinction among dark energy parameters may be artificial, but
it is sufficient for our goal to show the existence of some models which fit
the data set with a non-flat universe. Since the minimum values of $\chi^2$ 
for each model with fixed $z_*$ are smaller than the $\chi^2$ minimum for the case 
of cosmological constant, it is at least sensible to consider such model parametrization.
For example, the total $\chi^2$ minimum for cosmological constant is 177.99
when gold data is used for SNeIa while the model with $z_* = 0.5$ gives
$\chi^2$ minimum as 173.50. 

Now we look into Fig.~\ref{fig:OmOX} more closely.
For the cases with a small transition redshift as $z_* \sim 0.1$ and 
a large transition redshift like $z_* \sim 2.0$, the region around a flat universe is
favored  (Fig.~\ref{fig:OmOX} a, f). 
Since the equation of state approaches to 
the case with a constant equation of state $w_X = w_1$
as  $z_* \rightarrow 0$, the constraint for small $z_*$ is expected to be 
similar to the one obtained for a constant $w_X$.
This case has been investigated in Ref.~\cite{Ichikawa:2005nb} and found to favor the region 
closely around a flat universe, consistent with the result of $z_*=0.1$ here.  
Meanwhile, the model with $z_* = 2.0$ has a similar $w_X$ evolution to 
the case of Eq.~(\ref{eq:a}). This case has also been investigated 
in Ref.~\cite{Ichikawa:2005nb} and again shown to favor the region 
around a flat universe. Thus, the result for $z_* = 2.0$ in this paper 
is considered to be reasonable.

\begin{figure}[t]
\begin{center}
\scalebox{0.6}{\includegraphics{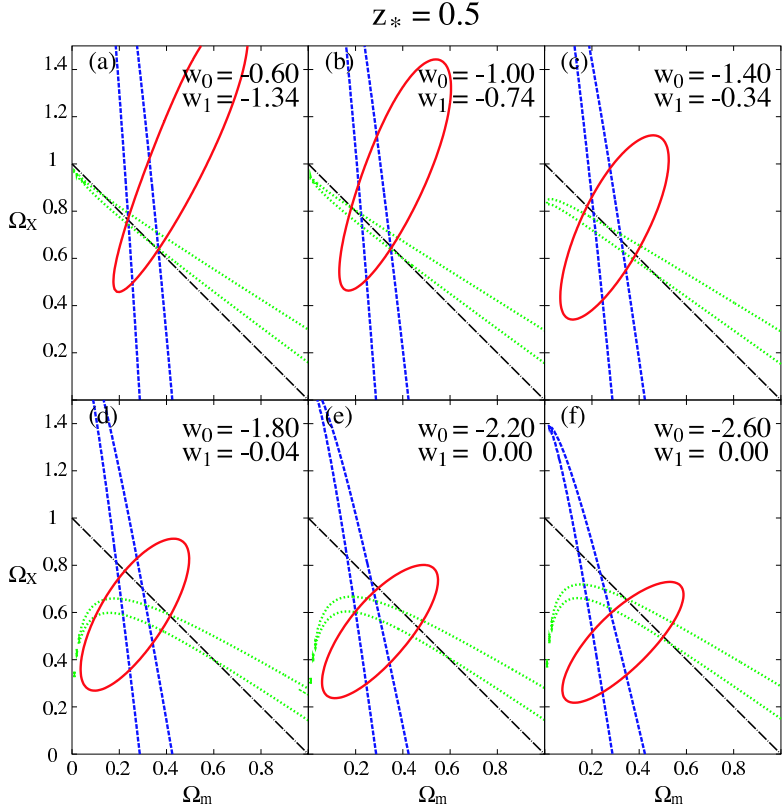}}
\caption{2$\sigma$ constraints on the $\Omega_m$ vs.  $\Omega_X$ plane
  from SNeIa (red solid line), BAO (blue dashed line), CMB shift
  parameter (green dotted line) are shown for the models with the transition
  redshift $z_* = 0.5$. $w_0$ is fixed to be 
  $-0.6$ (a), $-1.0$ (b), $-1.4$ (c), $-1.8$ (d), $-2.2$ (e) and $-2.6$ (f). 
  $w_1$ is taken to the value which gives minimum combined $\chi^2$ on each panel,
that is, $w_1= -1.34$ (a), $-0.74$ (b), $-0.34$ (c), $-0.04$ (d),
$0.00$ (e) and $0.00$ (f).  }
\label{fig:z0.5}
\end{center}
\end{figure}

\begin{figure}[t]
\begin{center}
\scalebox{0.6}{\includegraphics{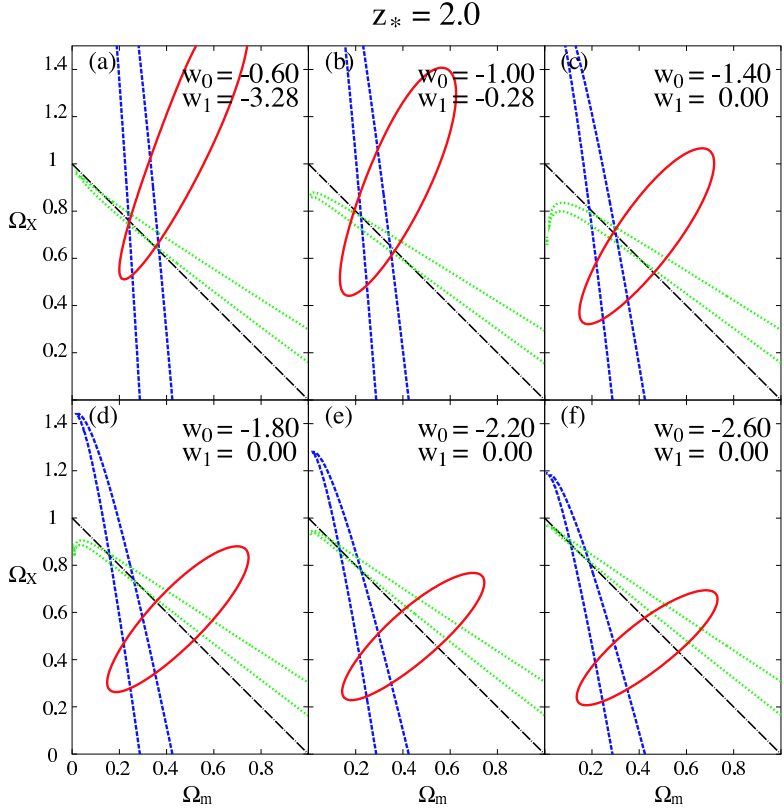}}
\caption{The same as Fig.~\ref{fig:z0.5} except that $z_* =2.0$. 
$w_1$ is taken to be $-3.28$ (a), $-0.28$ (b) and $0.0$ (c-f).}
\label{fig:z2.0}
\end{center}
\end{figure}

On the other hand, for the case with an intermediate transition redshift 
$0.5 \lesssim z_*\lesssim 1.0$, the
situation is different. We can see that the allowed region include a
relatively large region of an open universe (Fig.~\ref{fig:OmOX} b, c). 
 Since we marginalize
over $w_0$ and $w_1$, it is not easy to understand how this
happens. Thus, in Figs.~\ref{fig:z0.5} and \ref{fig:z2.0}, 
we also show the constraints from each observation separately 
fixing these parameters respectively with $z_*=0.5$ and 2.0. 
Here, gold data set is used for SNeIa data.
In these figures, we show the constraints on $\Omega_m$ and $\Omega_X$
 for the case with $w_0$ fixed to be 
  $-0.6$ (a), $-1.0$ (b), $-1.4$ (c), $-1.8$ (d), $-2.2$ (e) and $-2.6$ (f). 
For each $w_0$, we choose $w_1$ to give the minimum of total $\chi^2$. 
Then in Fig.~\ref{fig:z0.5} ($z_*=0.5$), 
$w_1=-1.34$ (a), $-0.74$ (b), $-0.34$ (c), $-0.04$ (d),
$0.00$ (e) and $0.00$ (f). 
In Fig.~\ref{fig:z2.0} ($z_*=2.0$), 
$w_1=-3.28$ (a), $-0.28$ (b) and $0.0$ (c-f).

A quick inspection of those figures reveals the difference of $z_*=0.5$ and 2.0
in Fig.~\ref{fig:OmOX} (b and f). In Fig.~\ref{fig:z0.5}, constraints from 
three observations cross at region around a flat universe with $\Omega_m \sim 0.3$
when $w_0 \gtrsim -1.5$ and $w_1 \lesssim -0.3$ (a-c). In addition, the observations allow
an open universe with $w_0=-1.8$ and $w_1 \sim 0$ (d), which corresponds to the elongated
feature of the contours in Fig.~\ref{fig:OmOX} (b). 
By contrast, in Fig.~\ref{fig:z2.0}, the observations only allow the region 
around a flat universe (a, b) and we do not find dark energy parameters 
such as to allow an open universe, as is shown in Fig.~\ref{fig:OmOX} (f). 

Since the contours of SNeIa and BAO have relatively similar degeneracy pattern 
with respect to the change of $w_0$ and do not depend much on the value of $w_1$
(these observations measure the distances to lower $z$ so they are controlled by $w_0$ 
rather than $w_1$),
we should look at CMB constraints to understand the difference between
Figs.~\ref{fig:z0.5} and \ref{fig:z2.0}. This is most noticeable on comparing panel (d) of 
the two figures.
In Fig.~\ref{fig:z0.5}, since the model has the transition redshft $z_*=0.5$ which is 
close to the present epoch, the value of $w_1$ mainly matters for the distance to 
the last scattering surface. Then the dark energy with $w_1 \sim 0$
leads to shorter comoving distance to the last scattering surface. 
This cancels the geometrical effect of an open universe 
and the angular diameter distance to the last scattering surface can have the observed value.
Such cancellation, however, does not seem to take place in Fig.~\ref{fig:z2.0} (d)
 where the transition redshft is as high as $z_*=2.0$ so the value of $w_0$ affects the acceleration. 
 With $w_0 = -1.8$, since the universe experiences significant amount of acceleration at late times,
 the observed angular diameter distance is not produced in an open universe. 
 From these considerations, we can conclude that the value of the transition redshft $z_*$
 plays important role in measuring the curvature of the universe. 
 
\begin{figure}[t]
\begin{center}
\scalebox{0.6}{\includegraphics{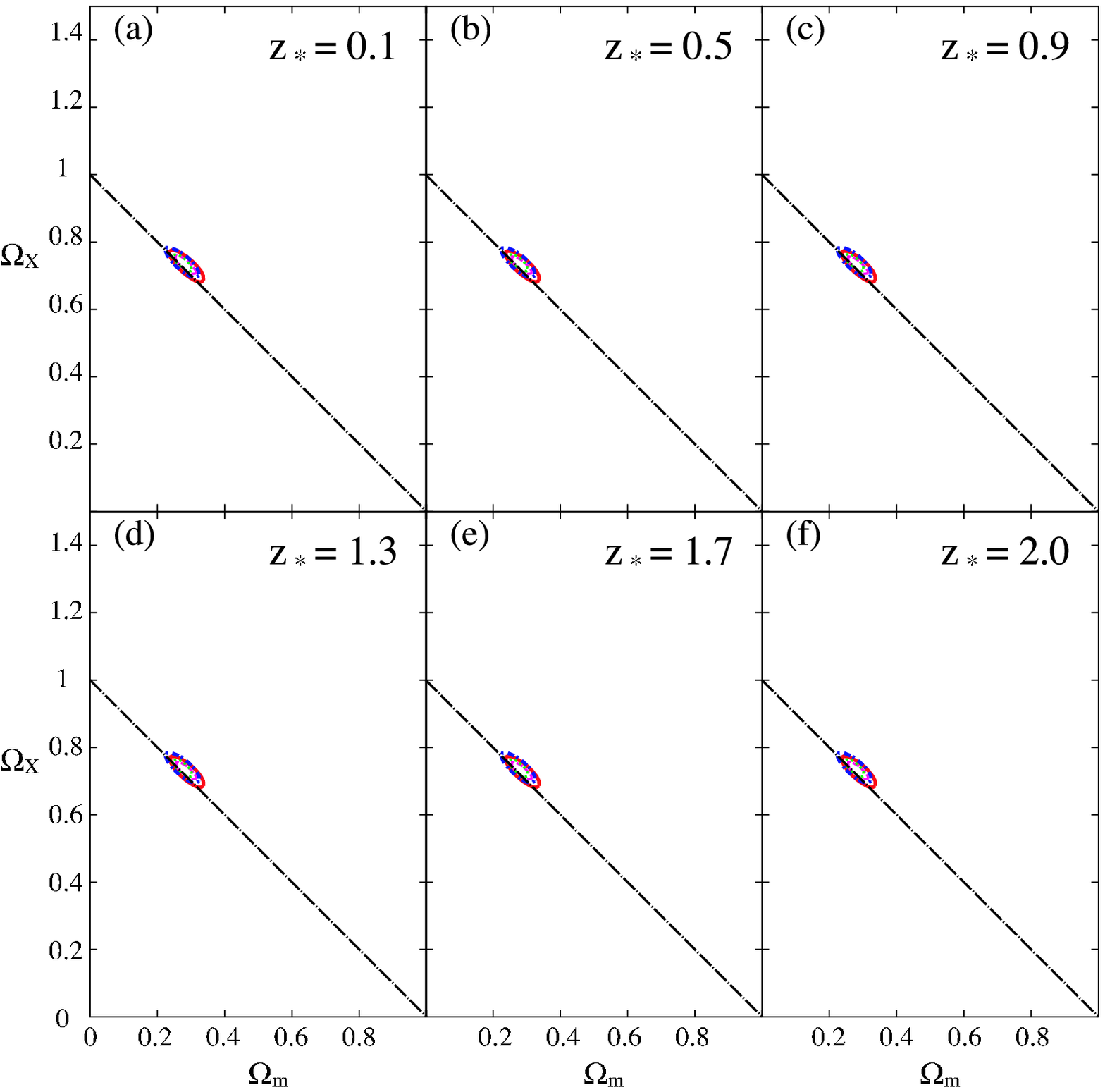}}
\caption{The same as Fig.~\ref{fig:OmOX} except that we assumed the
  prior $w_X \le -1$.  }
\label{fig:OmOX_phan}
\end{center}
\end{figure}

\begin{figure}[t]
\begin{center}
\scalebox{0.6}{\includegraphics{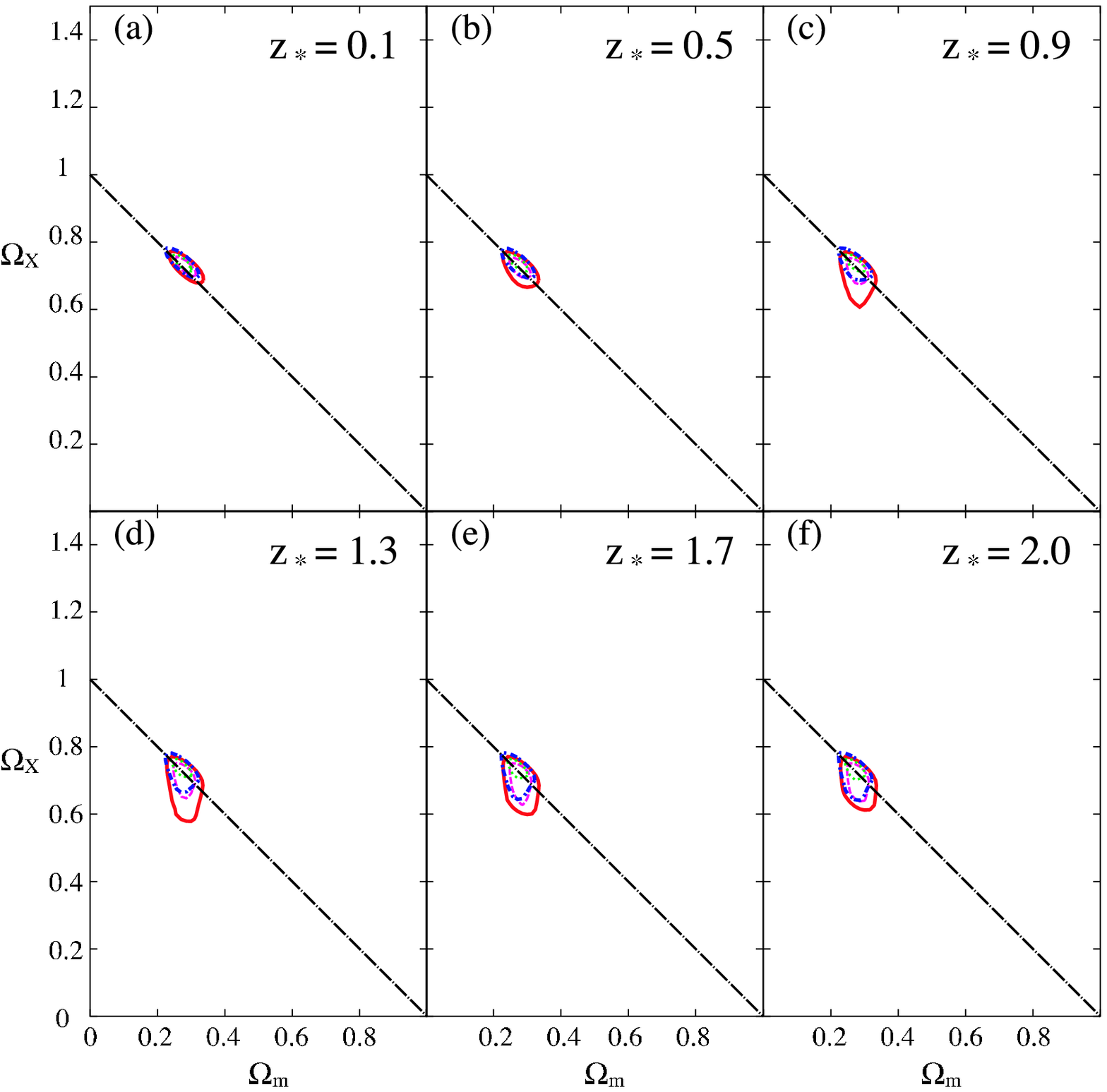}}
\caption{The same as Fig.~\ref{fig:OmOX} except that we assumed the
  prior $w_X \ge -1$.  }
\label{fig:OmOX_quin}
\end{center}
\end{figure}

From above discussion, we have noticed that, when an open universe is
allowed, the typical parameter range of the equation of state with the
parametriztion of Eq.~(\ref{eq:param}) is that $w_0 \le -1$, $w_1 \sim
0$ and $ 0.5 \lesssim z_* \lesssim 1.0$.  Since we need a large value
for $w_1$ such as $w_1 \sim 0$ for an open universe, it can be
expected that we would have a flat universe if we assume the prior on
the equation of state as $w_X \le -1$ (for the parameterization
adopted here, this also means that $w_1 \le -1$), i.e.,
so-called the phantom dark energy. To show this explicitly, we also
show the constraint on the $\Omega_m$ vs.  $\Omega_X$ plane with the
prior $w_X \le -1$ for several values of $z_*$ in
Fig.~\ref{fig:OmOX_phan}. We can see that the region around a flat
universe only is favored in this case. 
This is because $w_1$ can not be as large as 0 in phantom case by definition.
For reference, the constraint
for the case with $w_X \ge -1$ is also shown in
Fig.~\ref{fig:OmOX_quin}. In this case, although $w_0$ cannot be less
than $-1$, the values of $w_1$ can be large, which realizes the
situation where an open universe is favored to some extent. Thus the
allowed region slightly extends to the parameter space of an open
universe for some values of $z_*$.

Here we make some comments on the difference between the
parametrization adopted in this note and that of Eq.~(\ref{eq:a})
which was investigated in Ref.~\cite{Ichikawa:2005nb}. Although both
parametrizations accommodate the time dependence of a dark energy
equation of state, they are different in terms of how rapidly $w_X$
changes.  The parametrization we adopt here can rapidly change the value
of $w_X$ around the redshift where observations of SNeIa and BAO
mostly probe for suitable choice of the transition redshift. On the
other hand, for the models with Eq.~(\ref{eq:a}), the equation of
state changes its value more gradually compared to the parametrization
used here. Thus such cases is more or less similar to the case with a
constant equation of state, which is the reason why we got an almost
flat universe even if we allow $w_X$ to vary in time.  We have also
made the analysis for the models where the equation of state changes
with a step function at some transition redshift as 
$w_X  = w_0 \theta (z_* - z) + w_1 \theta (z- z_*)$. We obtained almost
the same tendency for the constraint as the parametrization of Eq.~(\ref{eq:param}).
 This means that models of dark
energy which have a rapidly changing equation of state around the
redshift range $ 0.5 \lesssim z_* \lesssim 1$, large region of an open universe
can be allowed. Thus
the assumption of a flat universe
may not be a valid assumption when one tries to constrain dark energy model
parameters for some particular models.

\section{Conclusion and discussion} \label{sec:conclusion}
We discussed how the nature of dark energy affects the determination
of the curvature of the universe.  In fact, it has been explicitly
shown that, when one adopts a dark energy model with a constant
equation of state or the parametrization with $w_X = \tilde{w}_0 + (1-a)\tilde{w}_1$,
current cosmological observations constrain the curvature of the
universe to be almost flat \cite{Ichikawa:2005nb}.  In this note, we
considered this issue using another parametrization which has the form
of Eq.~(\ref{eq:param}). We found that a flat universe is not always
favored for some particular cases.  As shown in Fig.~\ref{fig:OmOX},
when the transition redshift is in the range of $ 0.5 \lesssim z_*
\lesssim1$, the allowed region significantly extends to the parameter
space of an open universe.

We have also investigated what kind of dark energy can allow an open
universe by considering the constraint in detail. 
We have found that an open universe tends to be most favored 
when rapid transition such that $w_0 \le -1$, $w_1 \sim 0$ and $0.5 \lesssim z_*\lesssim1$
occurs. However, the period of $w_X < -1$ is not necessary for an open universe
to be allowed and actually we have shown  the allowed region extends to the space of an open
universe to some extent although it does not cover large region with the prior of
conventional dark energy: $w_X(z) \ge -1$. 
Finally, when the phantom dark energy $w_X(z) \le -1$ 
is assumed, the allowed region in the $\Omega_m$ vs. $\Omega_X$ plane
lies around the parameter space of a flat universe. 
Therefore, the dark energy has to behave like matter at early times, $w_1 \sim 0$,
in order for an open universe to be allowed.

The result presented in this note may have much implication for the
strategies to constrain models of dark energy. Although a flat
universe is usually assumed when one studies the nature of dark
energy, we have noticed that, for some models of dark energy, an open
universe is more favored than a flat universe. Thus a flat universe
may not be a good assumption to constrain some particular models of dark energy. On the
other hand, the above statement also means that one should be careful
to assume a model of dark energy when one tries to determine the
curvature of the universe since the assumption may affect the
constraint on it.


\end{document}